\documentclass[a4paper, oneside, twocolumn, notitlepage, 10pt]{extarticle_ecoc}

\newcommand{\mbf}[1]{\mathbf{#1} }

\newcommand{\E}{{\mathbb{E}}}
\newcommand{\fig}[1]{Fig.~\ref{fig:#1}}
\newcommand{\eq}[1]{Eq.~(\ref{eq:#1})}

\usepackage{soul}
\usepackage{color}
\usepackage{framed, xcolor}
\usepackage{multicol}
\usepackage{mathtools,amssymb}
\usepackage{cuted}
\usepackage{graphicx, epstopdf}
\usepackage{subfigure}
\colorlet{shadecolor}{yellow}

\usepackage{ecoc}

\begin{document}
\selectlanguage{english}    


\title{Rate Adaptive Geometric Constellation Shaping Using Autoencoders and Many-To-One Mapping}%


\author{
    Metodi P. Yankov\textsuperscript{(1)}, Ognjen Jovanovic\textsuperscript{(1)}, Darko Zibar\textsuperscript{(1)}, Francesco Da Ros\textsuperscript{(1)}
}

\maketitle                  


\begin{strip}
 \begin{author_descr}
 
   \textsuperscript{(1)} Department of Electrical and Photonics Engineering, Technical University of Denmark, 2800 Kgs. Lyngby, Denmark
   \textcolor{blue}{\uline{meya, ognjo, dazi, fdro@dtu.dk}}
   
 \end{author_descr}
\end{strip}

\setstretch{1.1}
\renewcommand\footnotemark{}
\renewcommand\footnoterule{}
\let\thefootnote\relax\footnotetext{978-0-7381-4679-9/21/\$31.00 \textcopyright 2021 IEEE}

\begin{strip}
  \begin{ecoc_abstract}
    A many-to-one mapping geometric constellation shaping scheme is proposed with a fixed modulation format, fixed FEC engine and rate adaptation with an arbitrarily small step. An autoencoder is used to optimize the labelings and constellation points' positions. 
  \end{ecoc_abstract}
\end{strip}


\section{Introduction}
With the exponentially growing data rate demands, optical communication networks are pressed to increase their efficiency. That implies that optical transceivers need to be able to take advantage of the available signal to noise ratio (SNR) in various and dynamic scenarios. In order to do so, the transceivers need to be rate adaptive. At the same time, efficiency requires that they operate as close as possible to the Shannon limit for the given effective SNR, which requires non-uniform distributed constellation set. Probabilistic amplitude shaping (PAS) \cite{Bocherer} has emerged as a very efficient method for achieving both of these requirements. However, it requires a distribution matcher, which is problematic to implement optimally and without a rate loss at very high speeds due to the requirement for serial processing of long bit sequences (1000s of bits \cite{Schulte}). It should be noted that approximate solutions exist that are able to achieve the required rate adaptivity at a slight rate loss, e.g. \cite{Yoshida, Yunus}. Nevertheless, geometric constellation shaping (GCS) is still of high interest as an alternative due to its potential to achieve shaping gain with conventional bit-interleaved coded modulation (BICM). The autoencoder (AE) concept has been applied successfully for optimization of GCS \cite{YankovECOC2022}, including in cases impaired by practical transceiver penalties. Rate adaptivity with GCS is typically challenging, since it requires to change the modulation format size and/or the forward error correction (FEC) overhead.  

In this paper, a GCS scheme is presented based on optimization using AEs. The scheme overcomes the challenges above and provides an extremely simple BICM system supporting rate adaptivity and achieving a shaping gain. This paper continuous the work in \cite{OgnjenCLEO}. Here, an arbitrarily small rate step is achieved by including the target rate into the optimization cost function. Furthermore, the benefits are demonstrated in a practical system employing practical FEC. 

\section{Proposed many-to-one system}
\label{sec:system}
\begin{figure}[!t]
\centering
 \includegraphics[trim=0 0 0 0cm, width=1.0\linewidth]{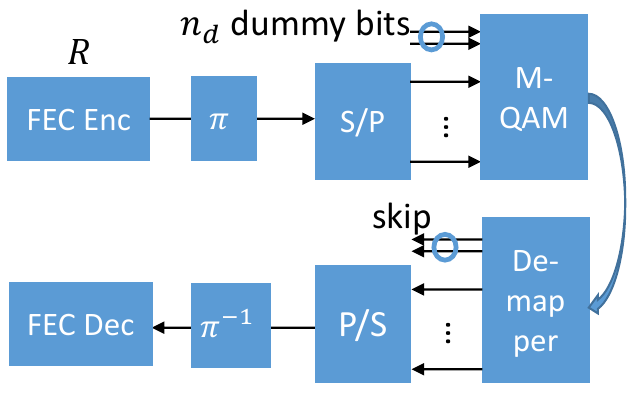}
 \caption{Block diagram of the proposed rate adaptive system.}
 \label{fig:MTO_scheme}
\end{figure}

The proposed method to achieve rate adaptivity is given in \fig{MTO_scheme}. It is a BICM system which is currently employed in most spectrally efficient communication standards, including coherent optical fiber systems. The only modification is the inclusion of $n_d$ dummy bits per symbol which do not carry data, but are assigned random bits. Their purpose is explained in the next section. The rate $R$ of the FEC and the size $M=2^m$ of the constellation (where $m$ is the length of the label) are fixed, which makes for a simple ASIC design. The net information rate (IR) that is targeted is $IR= R\cdot \left( m-n_d \right)$ bits/QAM symbol. The number $n_d$ may be fractional, which means that some bits of the label will carry both dummy and information bits. At the receiver, the dummy bits do not need to be demapped and can be skipped.  

\begin{figure*}[!t]
\centering
\subfigure[Complete constellation]{
 \includegraphics[trim=0 0 0 0cm, width=0.36\linewidth]{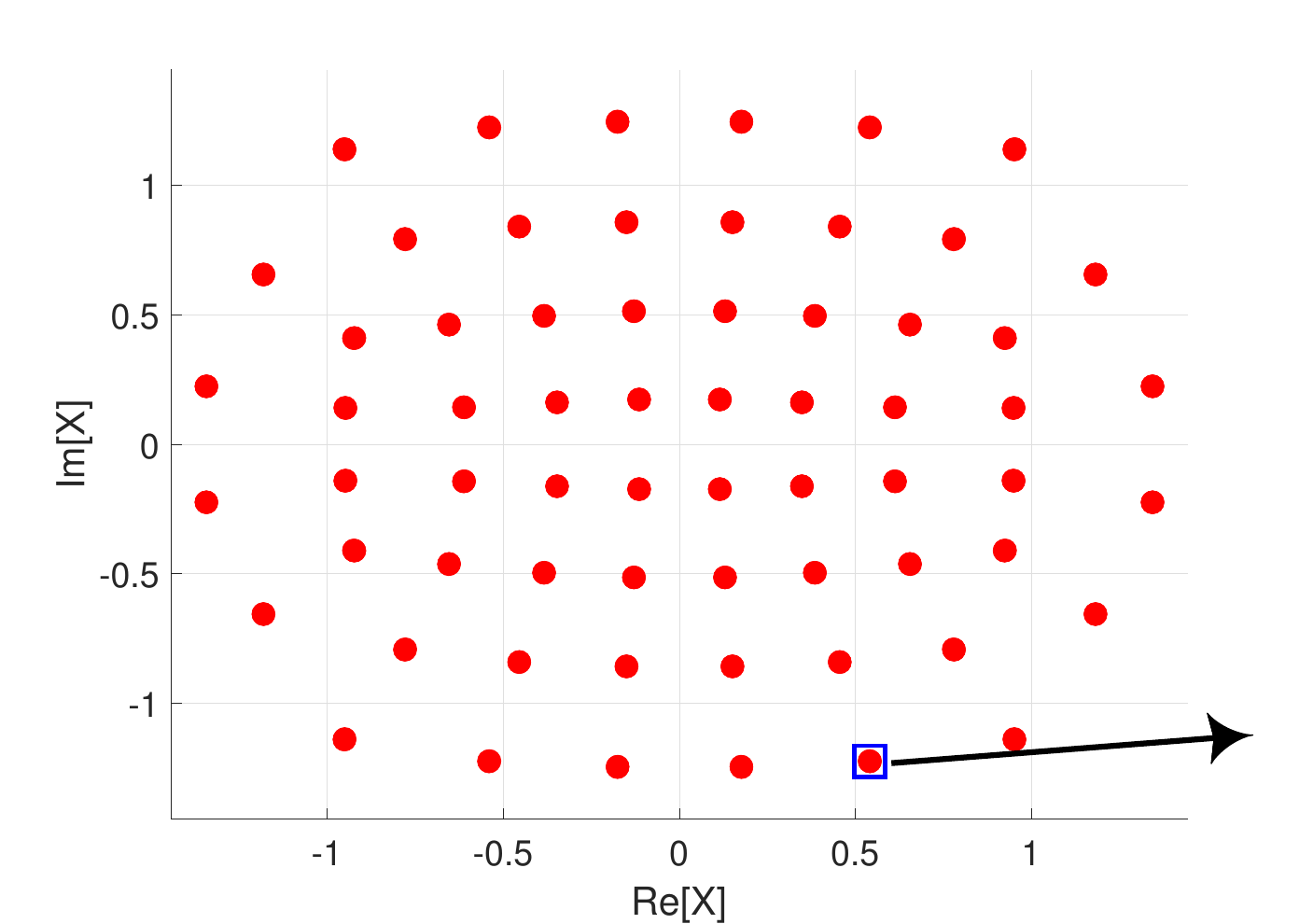}}
 \subfigure[Zoomed version]{
 \includegraphics[trim=0 0 0 0cm, width=0.36\linewidth]{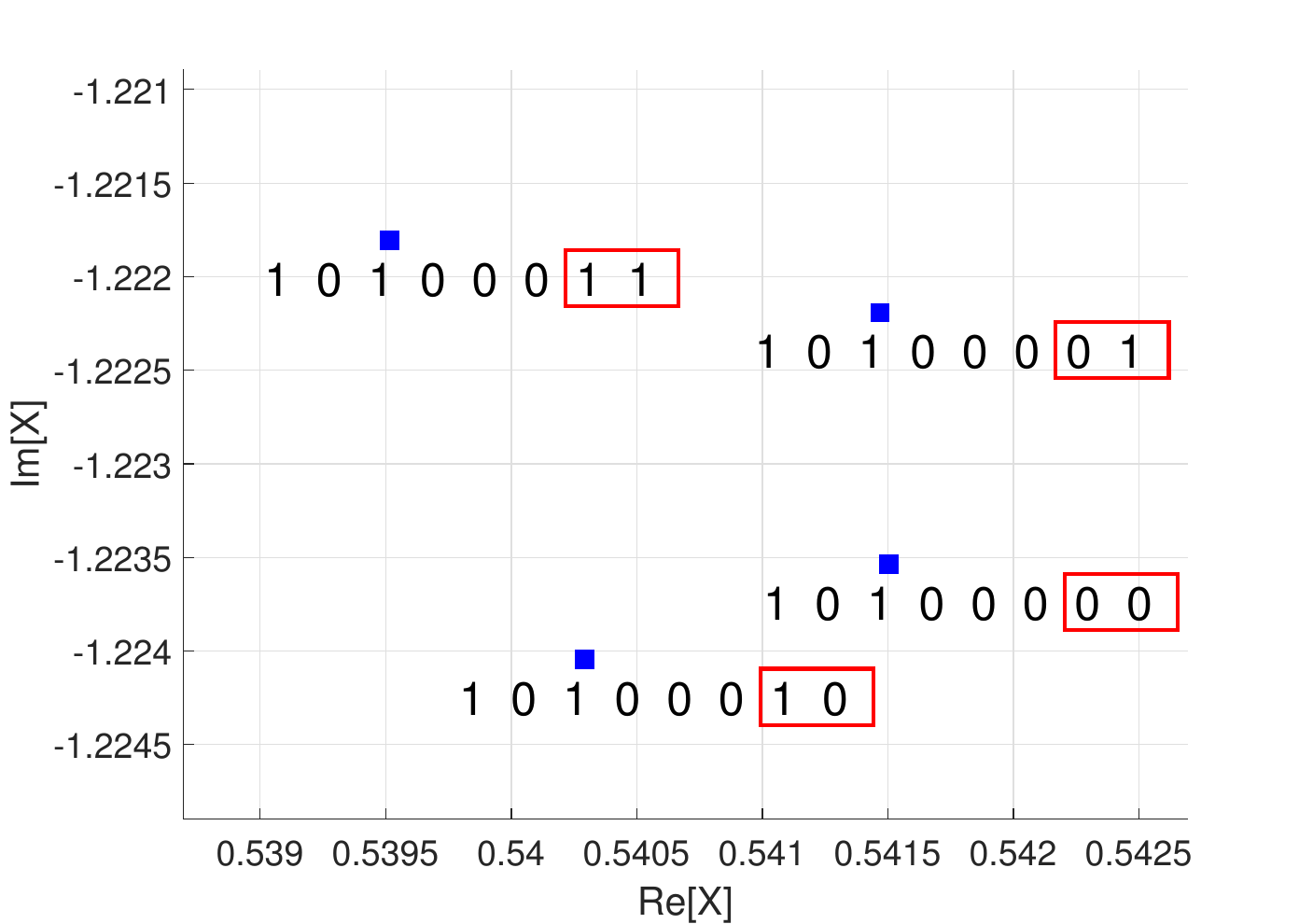} }
 \subfigure[Corresponding GMI]{
 \includegraphics[trim=0 -25 0 0cm, width=0.23\linewidth]{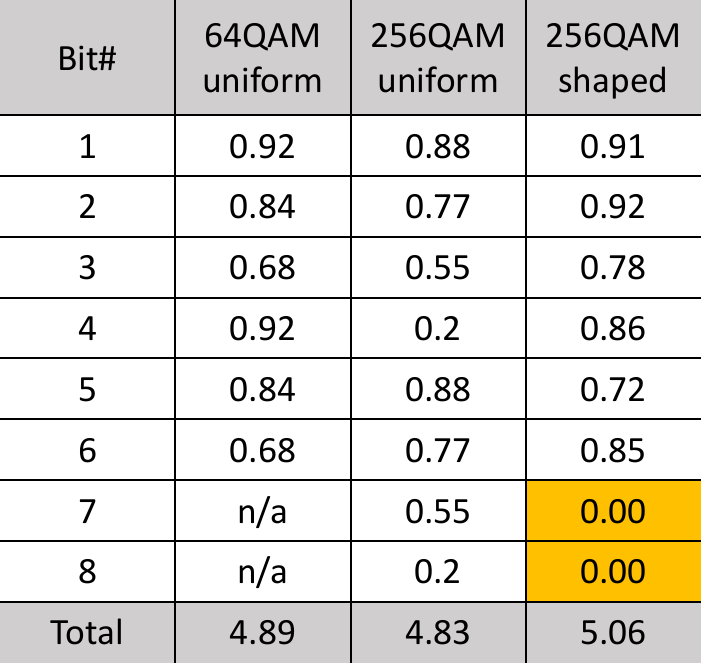} }
\caption{Optimized constellation for 15 spans of 100 km, 5 WDM channels, using the NLIN channel model, $m=8$. }
 \label{fig:constellation}
\end{figure*}

\section{Autoencoder-based optimization}
The AE concept is described in detail in e.g. \cite{YankovECOC2022} and references therein and its details are omitted here. In summary, the AE uses a neural network (NN) as encoder which maps sequences of bits (which we call 'labels') to constellation points, and a NN as decoder which produces approximations to the posterior probabilities of the bits. The weights of the NNs are optimized w.r.t. both the labeling and the points' positions to achieve a high generalized mutual information (GMI). For optimization, the nonlinear interference noise (NLIN) model \cite{NLIN} is used as a channel, which takes into account the contribution of the constellation moments to the NLIN. The encoder and decoder NNs have 2 hidden layers with 256 nodes each. The input to the encoder has a dimensionality of 8 bits, meaning that a constellation of 256 points is mapped uniquely.

The cost function used to support cases of $n_d > 0$ is given in \eq{cost}, where $k$ is time index, $\mbf{u}_k^i$ is the $i-$th bit in the label $\mbf{u}$, $p(\mbf{u}_k^i|y_k)$ is the corresponding posterior probability of that bit, $y_k$ is the channel output, and $\lfloor \cdot \rceil$ is the nearest integer operator. The term under the sum is the binary cross-entropy (CE) function per bit.  
 \begin{align}
 \label{eq:cost}
 Cost = -\frac{1}{m-\lfloor n_d \rceil} \sum_{i=1}^{m-\lfloor n_d \rceil} \E_k \left[ p(\mbf{u}_k^i|y_k) \right]
 \end{align}

The cost function disregards $\lfloor n_d \rceil$ bits of the label, and aims at improving the CE and thus the GMI of the remaining bits which will be used for data transfer. The result is a label, in which the $\lfloor n_d \rceil$ bits do not influence the location of the point on the I/Q plane. That means that bits which do not influence the cost can be assigned to symbols, which need not necessarily be distinguishable. Ideally, that results in a many to one mapping (MTOM) function. In practice, because we wish to maintain the size of the constellation, the mapping is unique, but those points are located at a very small Euclidean distance (ED), allowing for increased ED between sub-sets of constellation points which are selected with data-carrying bits. In summary, the GMI performance of the dummy bits is sacrificed in favor of improved ED between symbols with data-carrying bits. As such, the dummy bits are not 'wasted', but serve a very important function. It is worth noting that the family of regular QAM constellations of different size may be defined using this notion. For example, merging 4 nearest neighbors from a binary-reflected Gray coded (BRGC) 256QAM constellation and removing the least reliable bits from the label results in a BRGC 64QAM constellation. 


An example of the optimized constellation set for a system with 15x100 km spans, 5 WDM channels, 32 GBd at 50 GHz spacing and $n_d = 2$ is given in \fig{constellation} \textbf{a)}. The resulting effective received SNR is $\approx$ 15.75 dB, $\approx$ 15.76 dB and $\approx$ 15.99 dB for regular 64QAM, regular 256QAM and shaped 256QAM, respectively, at the respective optimal launch power. The constellation points are grouped in sub-sets of $2^{n_d}=4$. In each subset, the first $m-n_d$ bits in the label are unique, while the bits in the squares take any value. A zoom-in around one group of points is shown in \fig{constellation} \textbf{b)}. The achieved GMI per bit and the total GMI of the constellation are given in \fig{constellation} \textbf{c)}. 

\section{Results}
\begin{figure}[!t]
\centering
 \includegraphics[trim=0 0 0 0cm, width=1.0\linewidth]{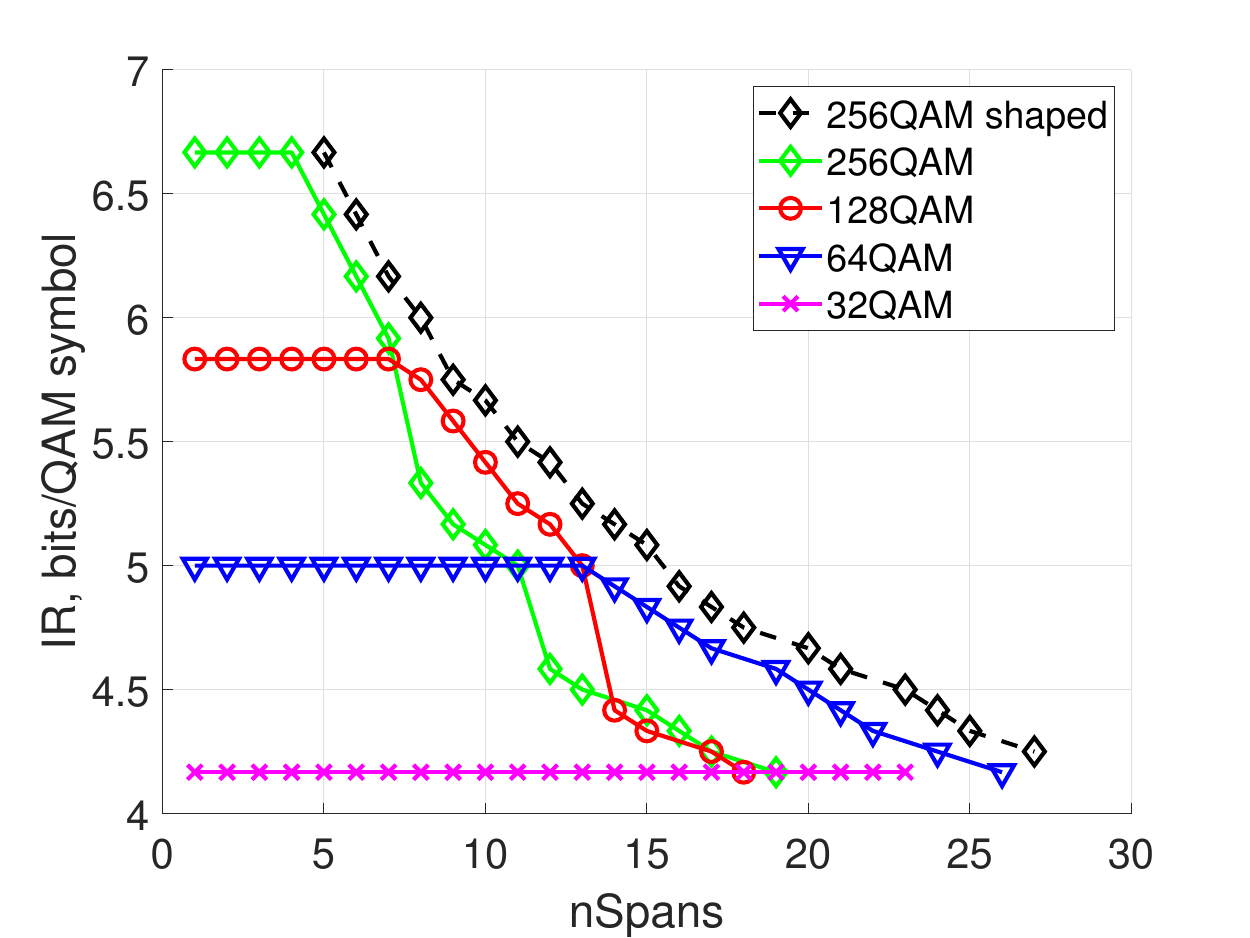}
\caption{Performance of the proposed rate adaptive scheme on a NLIN WDM channel assuming ideal FEC of $R=5/6$.}
\label{fig:performance_ideal}
\end{figure}

The maximum achievable data rate on the NLIN channel assuming an ideal FEC using this scheme may be estimated from the maximum distance, at which the GMI for the specific $n_d$ value is higher than the target $IR$ specified above. The target rate is swept in the range $\left[R\cdot 5; R \cdot 8 \right]$ by sweeping $n_d$ with a step of 0.1. In the case of the optimized MTOM labels, during testing, the encoder NN is replaced by a look-up table, and the decoder NN is replaced by an auxiliary Gaussian receiver~\cite{YankovJSTQE}. The achievable IR is given in \fig{performance_ideal} assuming $R=5/6$ (20\% overhead). The proposed system achieves a steady shaping gain of up to 2 spans (200 km) over all conventional modulation formats. The latter cannot be used with the proposed rate adaptive system for $n_d >= 1$. Instead, a switch of modulation format is needed. 

The MTOM labels with GCS are then tested using a more realistic model of optical fiber communications using a split-step Fourier method for solving the nonlinear Schr\"{o}dinger equation. We apply a similar 5-channel system as the one used for optimization, but transmit actual encoded data using a practical FEC (DVB-S2 low-density parity check (LDPC) \cite{DVBS2}). We also emulate frequency and phase noise (50 MHz offset, 10 kHz linewidth) impairments and apply the pilot-based DSP from \cite{YankovJSTQE} to detect the signal.

\begin{figure}[!t]
\centering
 \includegraphics[trim=0 0 0 0cm, width=1.0\linewidth]{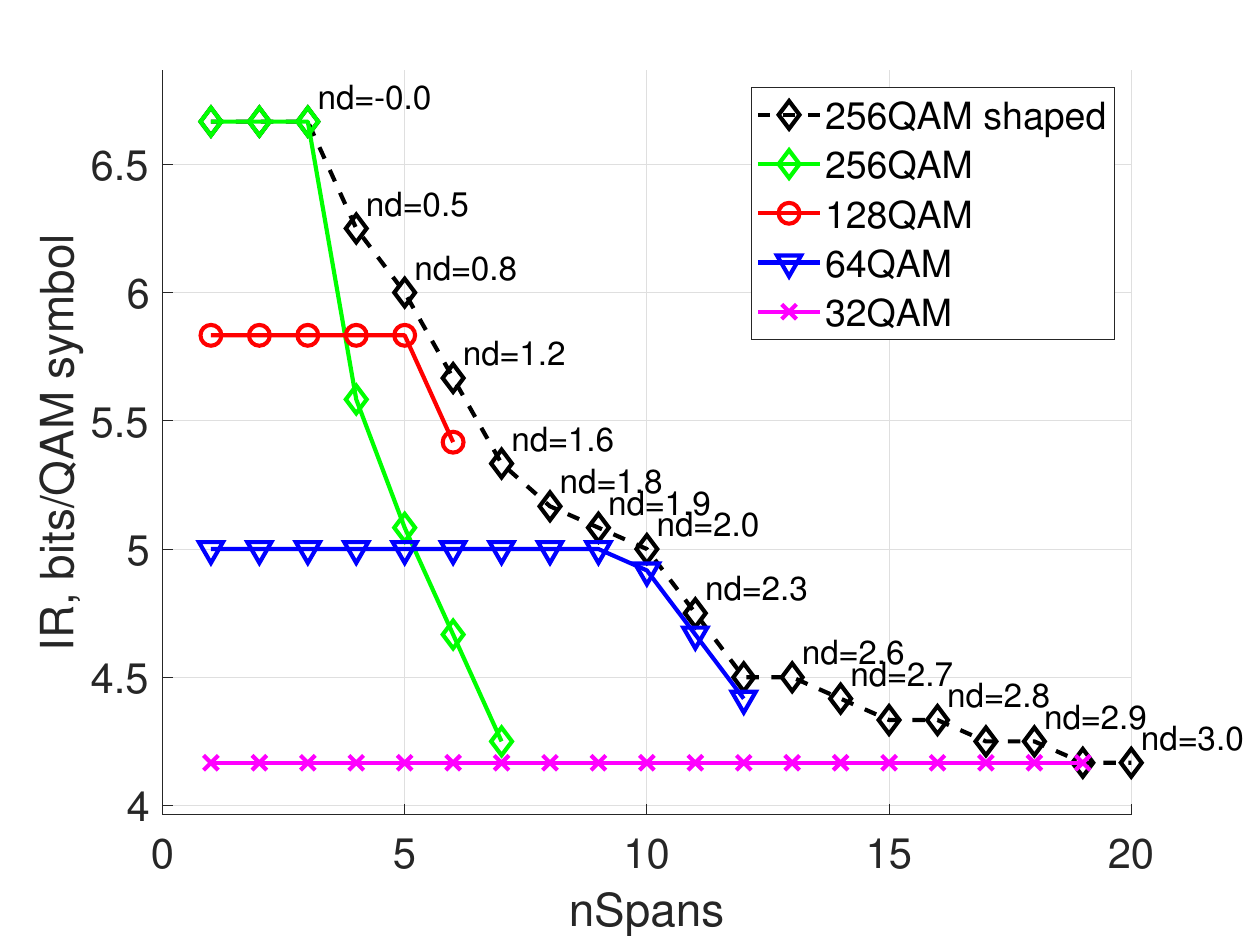}
\caption{Performance of the proposed rate adaptive scheme on a SSFM WDM channel with practical FEC of $R=5/6$ and practical receiver processing. }
 \label{fig:performance}
\end{figure}

More than $10^{6}$ information bits are simulated in each case. It is assumed that a target data rate is achieved without errors when the bit error rate after the LDPC decoder is below $5 \cdot 10^{-5}$, which is an example threshold of a potential outer hard-decision FEC\cite{Millar}. The maximum distance, at which the given target data rate is achieved error free is given in \fig{performance}. The annotations indicate the number $n_d$ which is used at the corresponding distance. The chosen span length of 100 km results in effective SNR jumps between spans that are too large to take full advantage of the fine granularity of the rate adaption. The additional loss of shaping gain w.r.t. \fig{performance_ideal} is attributed to the relatively poor FEC with a gap to Shannon of 1 dB, and is expected to disappear with state of art FEC.

The shaping gain w.r.t. regular QAM of flexible size is up to 1 span. However, the flexibility allows for significantly improved performance in between the regular QAM operating points, and even more w.r.t. the rate adaptive scheme with regular BRGC QAM. The latter quickly suffers severe performance degradation due to the penalty in GMI at the relatively low SNR w.r.t. QAM size. The proposed scheme exploits the increased ED between point groups of interest, and maintains stable performance. 

We point out that the cost function in \eq{cost} does not allow to specify indices in time which should be used for data at the ambiguous bits positions. This has the effect of \textit{puncturing} for $n_d < \lfloor n_d \rceil$, which is not always efficient depending on the chosen FEC. We therefore expect further improvements with properly selected FEC and carefully chosen encoded bits to be mapped to the fractional bit in the label. This challenge is left for future research.


\section{Conclusions}
A many-to-one mapping scheme was proposed for achieving rate adaptation using fixed FEC and modulation format size mapping/demapping engines. For each rate, many-to-one mapping functions were optimized using an autoencoder. The proposed system is directly integrable with conventional BICM and does not require any additional blocks. In fact, due to the fixed modulation format size, the mapping/demapping complexity is lower because it requires less chip area. The system is a solid candidate for next generation of coherent receivers.  

\section{Acknowledgements}
This work was supported by the Danish national research foundation SPOC project, ref. DNRF123, ERC-CoG FRECOM project ref. 771878, and the Villum Young Investigator OPTIC-AI, ref. 29334.


\printbibliography
\vspace{-4mm}

\end{document}